\documentclass{article}
\usepackage{graphicx,subfigure}
\usepackage{arxiv}
\usepackage[utf8]{inputenc} 
\usepackage[T1]{fontenc} 
\usepackage{hyperref} 
\usepackage{url} 
\usepackage{booktabs} 
\usepackage{amsfonts} 
\usepackage{nicefrac} 
\usepackage{microtype} 

\title{Hierarchical hidden community detectionr for protein complex prediction}

\author{
Chao Li \\
School of Computer Science and Technology\\
University of Science and Technology\\
Wuhan, 430074, China \\
\texttt{D201880880@hust.edu.cn} \\
\And
Kun He \thanks{Corresponding Author}\\
School of Computer Science and Technology\\
University of Science and Technology\\
Wuhan, 430074, China \\
\texttt{brooklet60@hust.edu.cn} \\
\And
Guangshuai Liu\\
School of Computer Science and Technology\\
University of Science and Technology\\
Wuhan, 430074, China \\
\And
John E. Hopcroft\\
Department of Computer Science\\ 
Cornell University\\ 
Ithaca NY, 14853, U.S.A.\\
}

\begin{document}
\maketitle

\begin{abstract}
\textbf{Motivation:} Discovering functional modules in protein-protein interaction networks through optimization remains a longstanding challenge in Biology. Traditional algorithms simply consider strong protein complexes that can be found in the original network by optimizing some metric, which may cause obstacles for the discovery of weak and hidden complexes that are overshadowed by strong complexes. Additionally, protein complexes not only have different densities but also a various range of scales, making them extremely difficult to be detected. We address these issues and propose a hierarchical hidden community detection approach to accurately predict protein complexes of various strengths and scales.\\
\textbf{Results:} We propose a meta-method called HirHide (Hierarchical Hidden Community Detection), which can adopt any standard community detection method as the base algorithm and enable it to discover hierarchical hidden communities as well as boosting the detection on hierarchical strong communities. To our knowledge, this is the first combination of hierarchical structure with hidden structure, which provides a new perspective for finding protein complexes of various strengths and scales. We compare the performance of several standard community detection methods with their HirHide versions. Experimental results show that the HirHide versions achieve better performance and sometimes even significantly outperform the baselines.\\
\textbf{Availability:} All data and code are available at https://github.com/JHL-HUST/HirHide/\\
\textbf{Contact:} {brooklet60@hust.edu.cn}\\
\textbf{Supplementary information:} Supplementary data are available at \textit{Bioinformatics}
online.
\end{abstract}

\section{Introduction}

A protein complex is a group of proteins that interact with each other for some specific biological activities \cite{fiannaca2013knowledge}. The identification of protein complexes is crucial for predicting protein functions \cite{schwikowski2000network, king2004protein, xie2011construction, winterhalter2014pepper}, disease genes \cite{lage2007human, yang2011inferring}, phenotypic effects of genetic mutations \cite{fraser2007using}, and drug-disease associations \cite{yu2015inferring}. Given a protein-protein interaction (PPI) network, where nodes represent proteins and edges represent interactions, the protein complexes can be searched by detecting densely connected subgraphs in the network. Mathematically, such subgraphs are called communities, in which nodes are joined together in tightly-knit groups, and there are only looser inter-connections among the groups \cite{girvan2002community,newman2003structure}.\par

Community detection plays a significant role in biological network analysis and provides insight into the underlying structure existing in networks \cite{girvan2002community,newman2003structure}. Over the last decade, numerous algorithms have been proposed for detecting communities in social as well as biological networks. Early works focus primarily on identifying disjoint communities that partition the nodes~\cite{blondel2008fast,pons2005computing,rosvall2008maps}, then researchers observed the overlapping membership among the communities and develop overlap community detection techniques~\cite{ahn2010link,coscia2012demon, lancichinetti2011finding,yang2012community}. Some partition-based community detection algorithms are also extended to address the overlapping case \cite{gregory2009finding,zhang2007identification}. However, almost all these algorithms focus primarily on the prominent community structure in networks, which is potentially problematic. Weak communities that are shielded by some dominant communities are sometimes of high value. For example, real-world protein complexes are not always dense, sometimes they can be very sparse \cite{liu2016using}. As proteins typically get involved in several interactions, there exist many overlaps among the protein complexes, and thus weak protein complexes are usually hidden behind the stronger ones. Additionally, there could be some undiscovered protein interactions~\cite{stumpf2005subnets}, thus we have no connection between these proteins in PPI networks. As a consequence, sparse protein complexes may be only sparse in existing incomplete PPI networks. But they are more likely to be dominant in real-world PPI networks if all protein interactions had been discovered. That is, sparse protein complexes may also be potential strong protein complexes. However, these sparse complexes are overlooked by previous technique.\par

Though standard community detection methods can find a portion of the sparse protein complexes by detecting weak communities, they cannot deal with the case where most nodes of the weak communities also belong to other stronger communities. Under such a case, these weak communities are defined as hidden communities~\cite{he2018hidden, he2015revealing}. In PPI networks, we call these communities the hidden protein complexes. For instance, in Fig. \ref{hierarchical_hidden} (a), we build a network with hidden community structures. Because standard algorithms focus on discovering dominant communities, the weaker community with green nodes is generally overlooked. Even if it is detected, its structure is considered as the structure of green nodes in Fig. \ref{hierarchical_hidden} (a), which contains four smaller blocks with dense intra-connections. However, its structure is actually like the structure of green nodes in Fig. \ref{hierarchical_hidden} (b). The community in (b) becomes detectable because the stronger communities have been weakened. And as edges belonging to stronger communities are removed, this hidden community doesn't contain four smaller blocks with dense intra-connections like (a) shows. Mathematical graph representations of many protein complexes related to sparser communities are hidden communities. They are partially or totally covered by stronger protein complexes, and finding such hidden complexes is very difficult. Traditional algorithms simply consider hidden protein complexes as a part of the stronger protein complexes, which causes big obstacles for the hidden ones to be discovered. This can partially explain why standard methods are not working well.\par

To address this problem, we design HirHide, which is inspired by a meta-approach called HiCode (Hidden Community Detection) \cite{HeSCHH15, he2018hidden}, a novel approach that first addresses the hidden community structure. However, HiCode does not consider hierarchical community structure, and hence it cannot handle complicated networks where communities are organized hierarchically, which exists in many real-world networks. In large-scale PPI networks, many protein complexes are organized hierarchically, indicating that protein complexes may consist of sub-complexes extending to several hierarchical levels deep. An example of such deeply embedded complex is the SAGA complex (MIPS identifier 510.190.10.20.10), a multi-functional coactivator that regulates the transcription by RNA polymerase II \cite{nepusz2012detecting}.\par

The main contributions of this work include:
\begin{itemize}
\item We propose HirHide that combines the detection of hierarchical structure with hidden structure, and could detect communities of various strengths (related to density) as well as communities of various scales (related to size). 
\item HirHide is designed as a general method that can be combined with standard community detection methods and enable them to discover hierarchical hidden communities and boost the detection of hierarchical dominant communities.
\end{itemize}

Throughout our experiments, level is used to denote the stratification of hierarchical structure and layer is used to denote the stratification of hidden structure. Multi-level hierarchical structure together with multi-layer hidden structure build a multi-granular characteristic of HirHide (see Fig. \ref{hierarchical_hidden} (c)). In experiments, we set two layers for hidden structure (one is dominant, the other is hidden), two levels for hierarchical structures to avoid over-complication (a total of 4 granular divisions). Layer1-level1 represents the first level of layer 1, which corresponds to the root, strong communities; Layer1-level2 indicates the second level of layer 1, which is a more detailed division of the root, strong communities; Layer2-level1 indicates the root, hidden communities in layer 2; Layer2-level2 denotes a more detailed division of the root, hidden communities. In the following, the four community divisions are shortly recorded as $L_{11}$, $L_{12}$, $L_{21}$, and $L_{22}$. \par



\section{Methods}

\subsection* {Hiddenness Value}
A PPI network can be represented as a graph $G = (V, E)$, where $V$ is the node set and $E$ is the edge set. Suppose the network is divided into $K$ communities, denoted by $\mathcal{C}=\{C_1,...,C_k,...,C_K\}$. \par
Let $\mathcal{S}_k$ be the set of communities stronger than a community $C_k$:
\begin{equation}
\mathcal{S}_k=\{C_i | F_i > F_k,C_i \in \mathcal{C} \}.
\end{equation}
$F_k$ represents the strength of community $C_k$, which can be calculated by the modularity metric~\cite{newman2003structure}. The larger the value is, the stronger the corresponding community structure is. Modularity $Q$ is defined as:
\begin{equation}
Q=\frac{1}{2m}\sum_{i,j} \left[ A_{ij}-\frac{d_id_j}{2m} \right] \delta(Com_i,Com_j),
\end{equation}
where $m$ indicates the number of edges in the graph, $d_i$, $d_j$ separately indicate the degree of node $i$ and node $j$, $Com_i$, $Com_j$ separately indicate the community to which nodes $i$ and $j$ belong. And $\delta (Com_i,Com_j)$ indicates whether nodes $i$ and $j$ are in the same community. If so, $\delta (Com_i,Com_j)$ = 1, and otherwise 0. \par

\begin{figure*}[!tpb]
\centering
\subfigure[]{\label{fig:a}\includegraphics[width=0.28\textwidth]{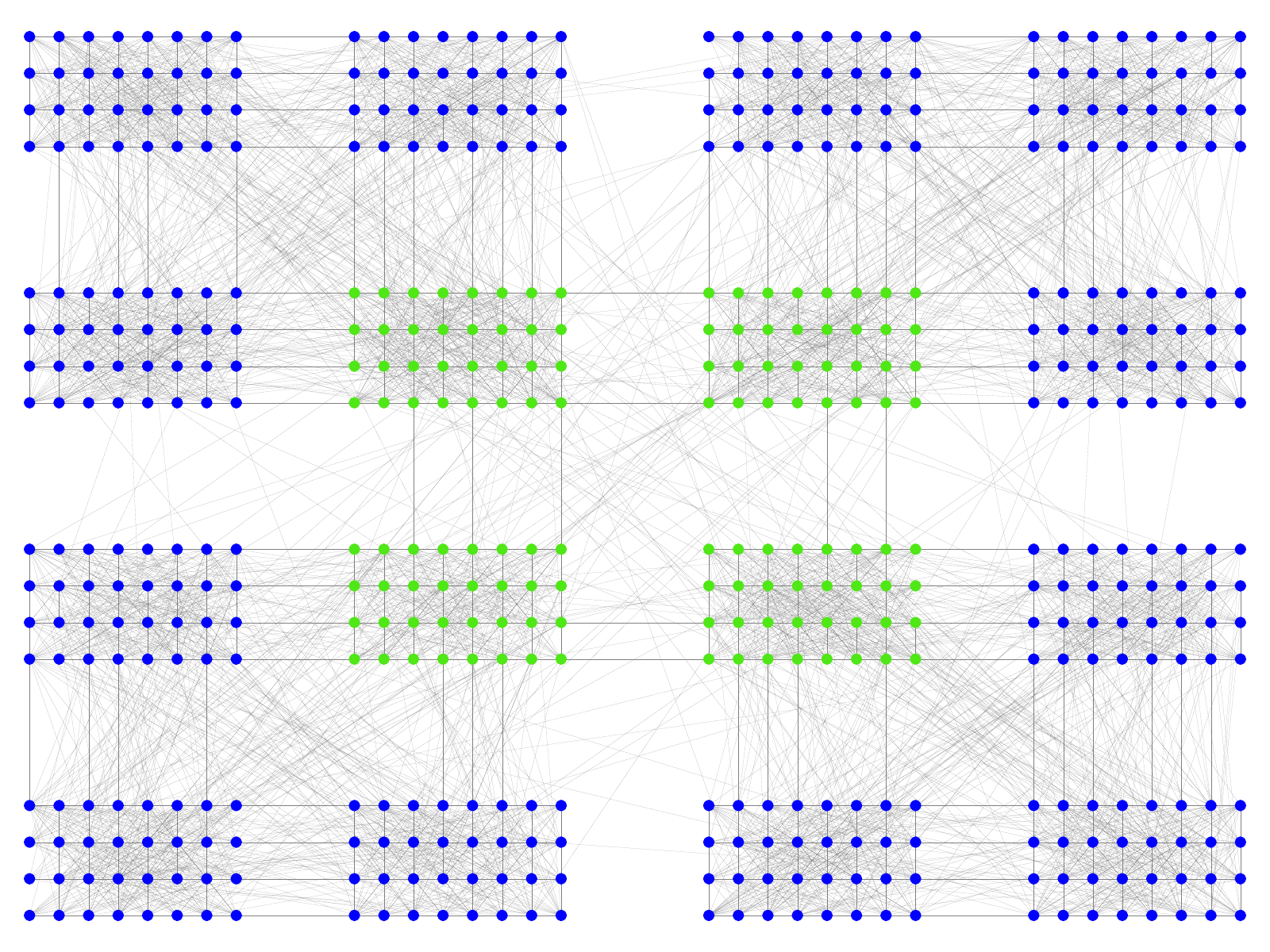}}
\subfigure[]{\label{fig:b}\includegraphics[width=0.28\textwidth]{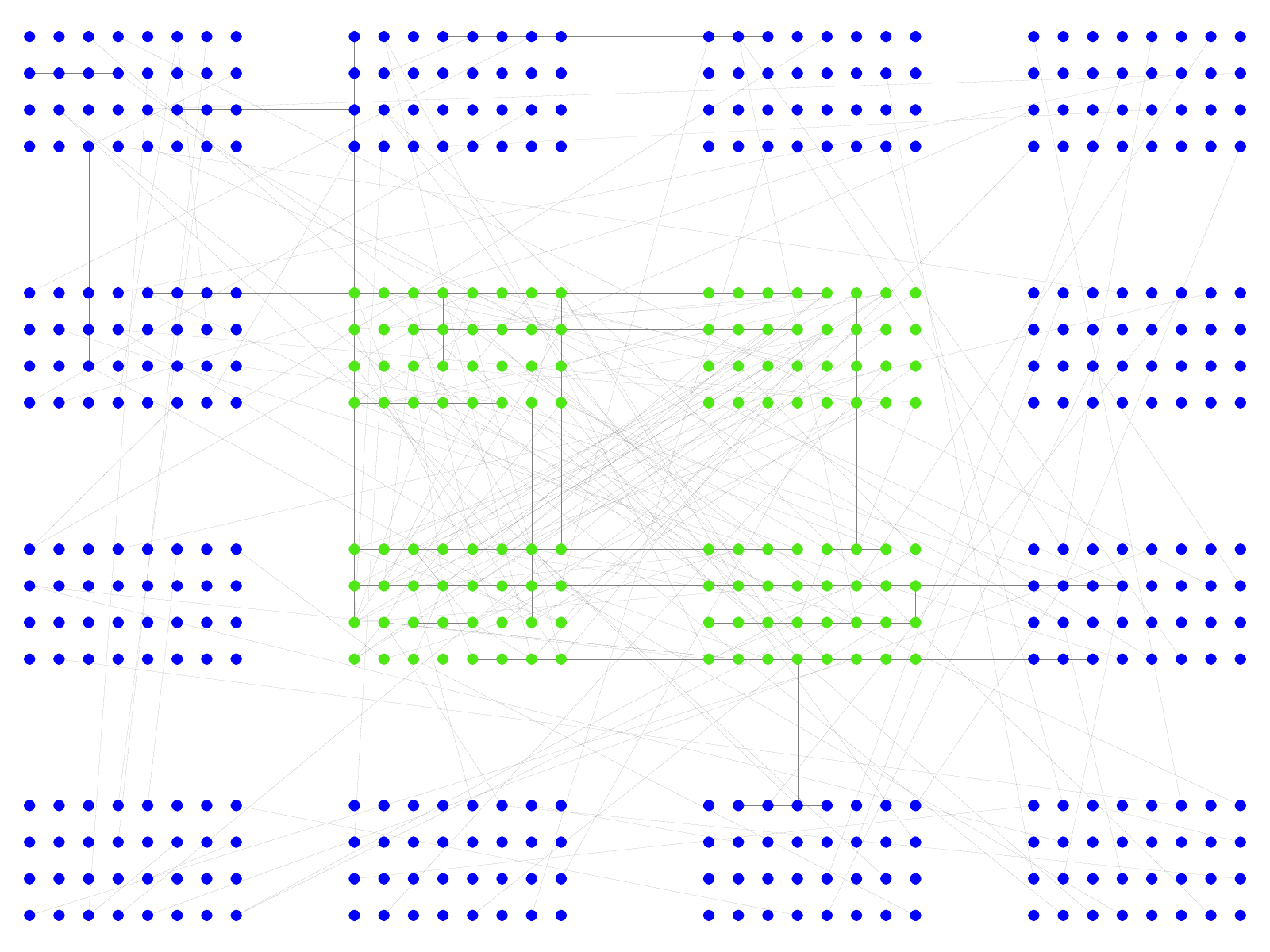}}
\subfigure[]{\label{fig:b}\includegraphics[width=0.42\textwidth]{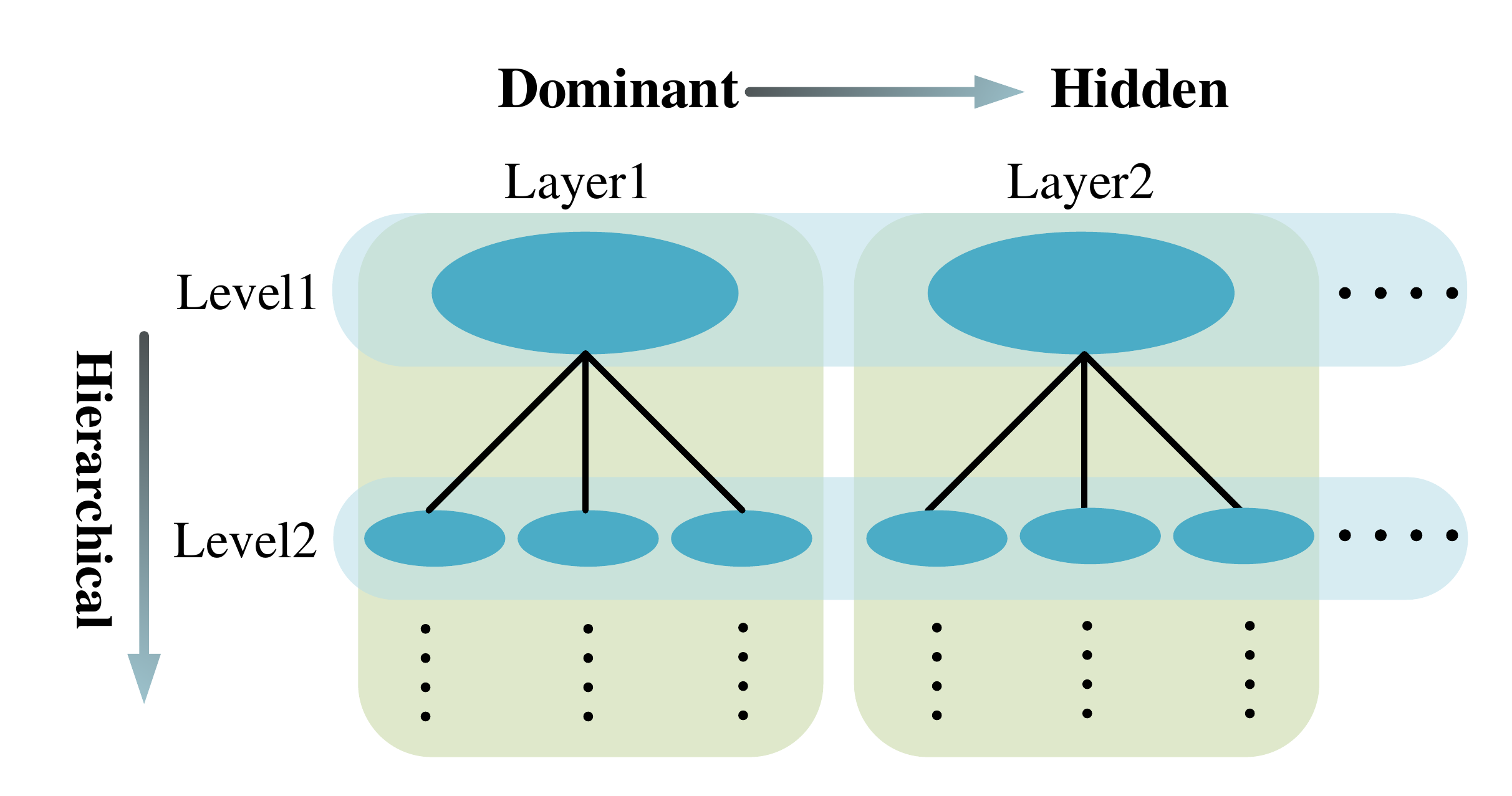}}
\caption{Example to show the hierarchical hidden structure. (a) A network with hierarchical hidden structure. The graph exists of four large dense communities consisting of 128 nodes, each with an internal subdivision of four small communities with 32 nodes. Additional, there is a sparse, hidden community structure consisted by the four small blocks in the center. (b) The network after weakening the stronger communities. The community with green nodes is the hidden community, detected by weakening the stronger communities. So it doesn't contain dense edges. (c) Schematic diagram of a hierarchical hidden structure. Level and layer are separately used to denote the stratification of hierarchical structure and hidden structure. Each level of hierarchical structure can contain several layers of hidden structure, and vice versa.}
\label{hierarchical_hidden}
\end{figure*}

He \textit{et al.} raise a formula to calculate the hiddenness value of a community \cite{he2018hidden}. This definition calculates the fraction of nodes of $C_k$ belonging to other stronger communities. However, this formula does not consider the structure information of $C_k$. A sparsely connected community should be more likely to be hidden than a densely connected community even if it has the same fraction of nodes belonging to other stronger communities. Consequently, we improve the concept of hiddenness value of a community $C_k$ as follows:
\begin{equation}
H(C_k)=\frac{\left|\bigcup_{C_i \in \mathcal{S}_k} C_i\cap C_k \right|}{|C_k|\cdot}\cdot \frac{1}{\sigma(Q_k)}.
\end{equation}
The first term is the fraction of nodes in other stronger communities. The second term indicates how hidden community $C_k$ could be: $Q_k$ is the modularity of $C_k$ ranging from [-0.5,1], and $\sigma$ represents the sigmoid function that can proportionally shift $Q_k$ to a positive number. 
The larger the hiddenness value is, the higher the probability the community is hidden, and the less likely the community can be detected. Note that there is no single specific threshold between a hidden community and a dominant community. For two communities ${C_i,C_j} \in \mathcal{C}$, if $H(C_i) > H(C_j)$,
then $C_i$ is comparatively hidden compared to $C_j$, and $C_j$ is
comparatively dominant compared to $C_i$.

\subsection*{Hierarchical communities versus hidden communities}
Hierarchical community structure and hidden community structure are two aspects of community structure. An algorithm combining these two concepts may be confusing. For example, sometimes a subgraph with a hidden community and one or multiple stronger communities can be wrongly detected as a hierarchical structure. Fig. \ref{hierarchical_hidden} (a) and (b) show this case. So we first discuss the difference between hidden structures and hierarchical structures to clarify the concepts. When considering the hierarchical community structure, an algorithm gradually detects stronger and smaller communities, and ignore weaker communities. When considering the hidden community structure, an algorithm detects communities weaker than the overhead communities. Intuitively, when the density of a community is calculated by modularity, if the average density of root communities is 0.5, then the average density of the sub-hierarchical level is larger than 0.5 and the average density of the next hidden layer is less than 0.5. Note that, a hidden community can be totally covered by one stronger community or be partially covered by one or multiple stronger communities. And the hidden community does not need to be smaller than the overhead communities. Due to their different characteristics, each layer of hidden structure can contain several levels of hierarchical structure, and vice versa (Fig. \ref{hierarchical_hidden} (c)). HirHide combines the advantages of both types of algorithms, and guarantees a more complete protein complex detection.\par

\subsection*{HirHide}
We propose a hierarchical hidden detection approach (HirHide) for community mining tasks. Our algorithm consists of three steps. In the first step of initialization, HirHide identifies a layer of communities $Layer 1$ via the base algorithm (Fig. \ref{hierarchical_hidden} (c)), which can be a standard community detection algorithm with promising performance. \par

The second step is called the hierarchical detection step. HirHide constructs a hierarchical structure by recursively capturing sub-communities and iterating until an appropriate number of levels are found. A crucial dimension of this step is to determine the number of levels in a network. We simply count the number of nodes in each community. When the average number is smaller than a certain threshold, the algorithm stops capturing sub-communities. In our experiments, the default threshold is set to 9.\par

The third step is called the hidden detection step. HirHide weakens the structure of the previously detected layer $Layer_1$ to get a reduced graph $G'$. In $G'$, the base algorithm is used to detect new communities to form a hidden layer $Layer_2$. Because $G'$ does not contain the strong communities of $Layer_1$, the weaker communities can be easily discovered. 
Here the hierarchical detection step can be used on $Layer_2$ in the reduced graph $G'$ to find hierarchical structures for this layer. 
Then HirHide weakens $Layer_1$ and $Layer_2$ and calls the base algorithm again to detect $Layer_3$. This process iterates until no communities can be detected. \par

In HirHide, the step of digging out hidden structure is in a similar spirit with HiCode \cite{he2018hidden}. But the step of combining hidden structure and hierarchical structure is new and well-designed to guarantee both hierarchical communities and hidden communities to be well captured. A key issue is which level of hierarchical communities should be weakened in the detected layer. When these hierarchical levels are organized into trees with each community in the first level as root, we can weaken the communities in roots or the communities in leaves (Fig. \ref{hierarchical_hidden} (c)). Because the edge connection condition and the total size of nodes in each hierarchical level are similar (considering separate nodes and too small communities are removed) and the weakening step is a global operation, which level is weakened does not make a big difference. To double-check, we analyzed the performance of HirHide framework when it separately weakens the communities in roots or leaves in the confirmatory experiments.\par

In the hierarchical step, the sub-communities are captured on the original graph, but they can also be captured on the reduced graph $G'$ after weakening the structure of strong communities. Intuitively, the reduced graph has weakened the influence of other layers, so the detected sub-communities should be more precise. In our confirmatory experiments, the results of recursively capturing sub-communities on the original graph or the weakened graph are compared. \par

Although the base algorithm is recursively called in both the hierarchical step and the hidden step, its role is significantly different. In the hierarchical step, the base algorithm is repeatedly applied in sub-communities to discover smaller sub-communities. In the hidden step, the base algorithm is repeatedly used in the reduced graphs to dig out hidden communities. \par

\section{Experiments}

\subsection*{Algorithms, data, and metrics}
\subsubsection*{Algorithms}
We select three state-of-the-art algorithms as the baseline methods as well as the base algorithms of HirHide, which are MOD \cite{blondel2008fast}, Infomap \cite{rosvall2008maps}, LC \cite{ahn2010link}. None of them can detect hidden communities without HirHide. After they are combined with HirHide, their HirHide versions are called HirHide-MOD, HirHide-Infomap, and HirHide-LC. We also compare HirHide with the HiCode \cite{he2018hidden} version of the three methods, separately called HiCode-MOD, HiCode-Infomap and HiCode-LC. One of our metrics comes from ClusterONE \cite{nepusz2012detecting}, which does not provide the complete source code. As a consequence, we use it as a baseline method without the corresponding HirHide version. Overall, the algorithms involved in the comparisons are HirHide-MOD, HirHide-Infomap, HirHide-LC with HiCode-MOD, HiCode-Infomap, HiCode-LC, MOD, Infomap, LC and ClusterONE.\par

The above comparison involves four other algorithms. MOD \cite{blondel2008fast} treats each node as a separate community in initialization and gradually optimizes the modularity value by expanding the size of each community. After repeatedly iterating, MOD can get a community division with the largest modularity value. Infomap \cite{rosvall2008maps} is based on the principle of information theory and defines the community from the perspective of coding. To get the maximum compression ratio, Infomap uses Huffman coding and community structure secondary coding. In this way, the problem of community detection is transformed into an optimization problem: finding a community division so that the codeword length of random code walks within and between communities is the smallest. LC \cite{ahn2010link} reinvents communities as groups of links rather than nodes. This approach successfully reconciles the antagonistic organizing principles of overlapping communities and hierarchical structure. Link communities naturally incorporate overlap while revealing hierarchical organization. ClusterONE \cite{nepusz2012detecting} outlines the concept of cohesiveness score and uses a greedy growth process to find groups that are likely to correspond to protein complexes in a PPI network.

\subsubsection*{Data}
We compare these algorithms in three large scale yeast PPI networks, which consist of core experimental yeast PPI network \cite{krogan2006global}, a combined computational interaction network \cite{kim2013yeastnet} and the entire set of physical protein-protein interactions in yeast from BioGRID \cite{stark2006biogrid}. These three datasets are referred to as the $Krogan\_core$, $YeastNet$ and $BioGRID$. \par

To evaluate the performance of each algorithm, we use Munich Information Center of Protein Sequences (MIPS) \cite{mewes2004mips} and CYC \cite{pu2008up} as reference sets. The yeast protein complexes cataloged by the MIPS database have been widely used to generate protein-protein interaction reference sets. And CYC is a comprehensive catalog of manually curated 408 heteromeric protein complexes in S. cerevisiae. For convincing, we choose the latest version of the two reference sets. Additionally, we only consider complexes containing 3 to 100 proteins as the reference protein complexes to avoid the selection bias.\par

Table \ref{basic} shows the basic information of the three PPI networks. Because different PPI networks contain different nodes and edges and the two reference sets have different reference complexes, we have removed proteins that only exist in a PPI network or a reference dataset.\par

\begin{table}[!t]
\centering
\caption{Basic information of the three PPI networks.}
\footnotesize
\vspace{2pt}
\setlength{\tabcolsep}{3.6mm}
{\begin{tabular}{@{}lrrrr@{}}
\toprule \textbf{Dataset}& 
\multicolumn{2}{c}{\textbf{MIPS}}& \multicolumn{2}{c}{\textbf{CYC}}\\ 
& Nodes & Edges & Nodes & Edges \\\midrule
\textbf{ BioGRID }&1155&10825&1355&13098\\
\textbf{ YeastNet }&1085&11271&1289&13564\\
\textbf{ Krogan }&679&1797 &910&2808\\ \toprule
\end{tabular}}
\label{basic}
\end{table}

\subsection*{Evaluation Metrics}
It has become a standard practice to compare the performance of different methods by assessing their ability to identify the reference communities. To evaluate the performance of a community detection algorithm, the recognized metric is the F1 score \cite{chase2014thresholding}, which is the harmonic mean of the precision and recall. However, the reference communities, which only contain protein complexes whose interactions can be discovered under the current experimental conditions, have incomplete nature \cite{li2010computational}. Using the F1 score as an evaluation metric is unreasonable when the reference sets are incomplete. Because under the same condition, the more communities an algorithm detects, the smaller the precision is. So traditional algorithms are typically designed to detect fewer complexes than the reference complexes (much less than the real-world complexes) to increase the precision. Consequently, comparing the F1 score is unfair for the algorithms that detect more complexes.\par

As the current reference sets are incomplete \cite{li2010computational}, we evaluate the performance of the algorithms by two other measures. One is the maximum matching ratio (MMR), which is designed specifically for protein complex detection \cite{nepusz2012detecting}. 
MMR guarantees that each detected community only matches one reference community and vice versa. And it maximizes the total score of all one-to-one connections between predicted and reference complexes. This measure is inspired by the bipartite graph maximum matching problem, in which the two sets of nodes respectively represent detected complexes and reference complexes. MMR tries to find the best match for each reference complex. So even if an algorithm detects more complexes, MMR won't be reduced. \par

The other is the recall, which only measures the capacity of discovering the reference complexes. So additional hierarchical hidden communities do not decrease the score. The recall scores each pair composed of a predicted complex and a reference complex by their similarity. Given a set of detected communities $\mathcal{D}$ and a set of reference communities $\mathcal{G}$. Each reference community $G_j$ has its individual recall:
\begin{equation}
R(G_j)=\max \limits_{D_i\in \mathcal{D} } \frac{|G_j\cap D_i|}{|G_j\cup D_i|}.
\end{equation} 
The final recall is defined as the average of $R(G_j)$ over all reference communities.

\begin{figure*}[ht]
\begin{center}
\includegraphics[width=1\textwidth]{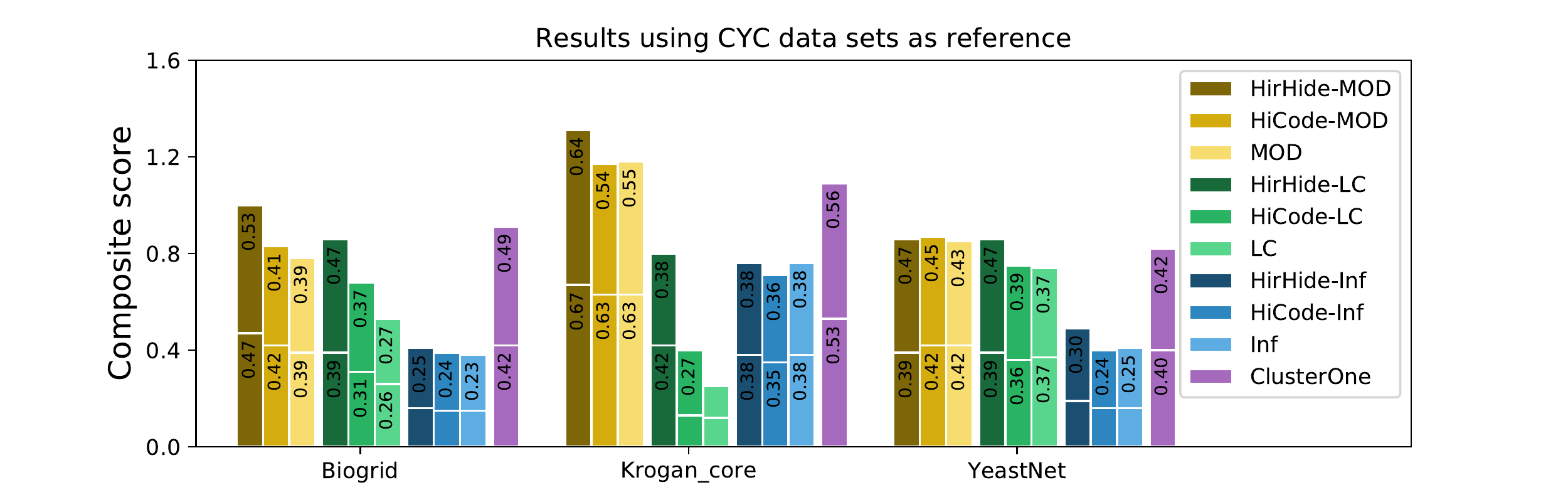}
\includegraphics[width=1\textwidth]{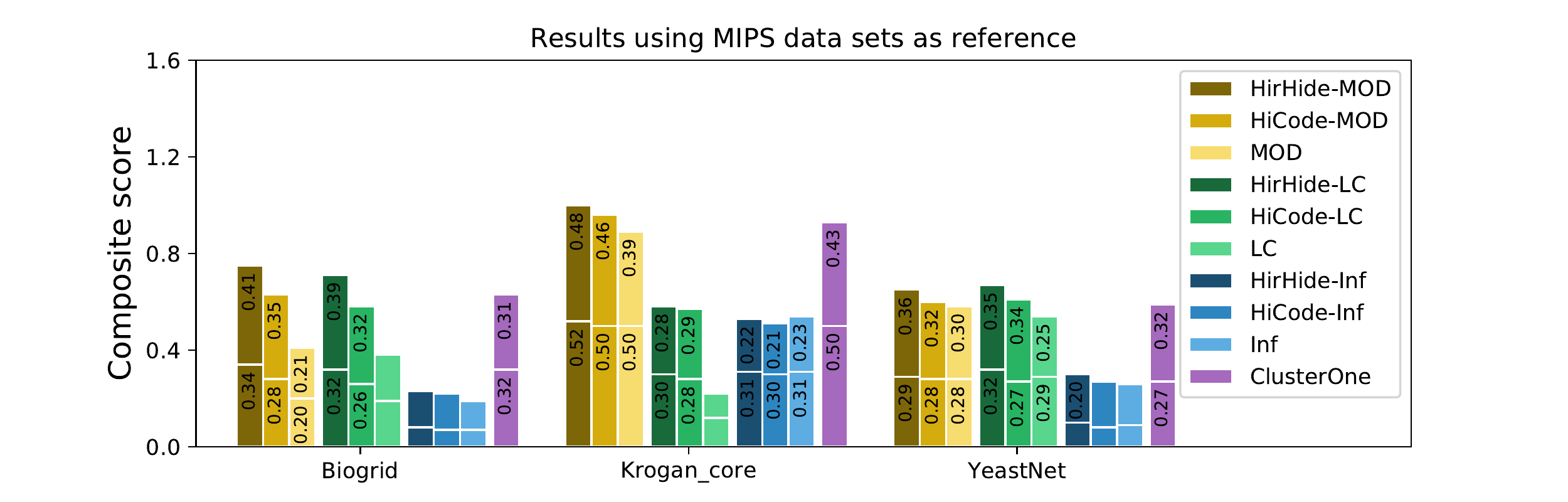}
\end{center}
\vspace{-1em}
\caption{Benchmark results. The different bar clusters represent results for different datasets ($Biogrid$, $Krogan\_core$ and $YeastNet$) as the PPI networks. The bars on bottom indicate the scores of maximum matching ratio (MMR) and the bars on top represent the scores of the recall. The number inside each bar is the score and higher bars represent better performance.
}
\label{benchmark reuslts}
\end{figure*}

\begin{figure*}[ht]
\begin{center}
\includegraphics[width=0.48\textwidth]{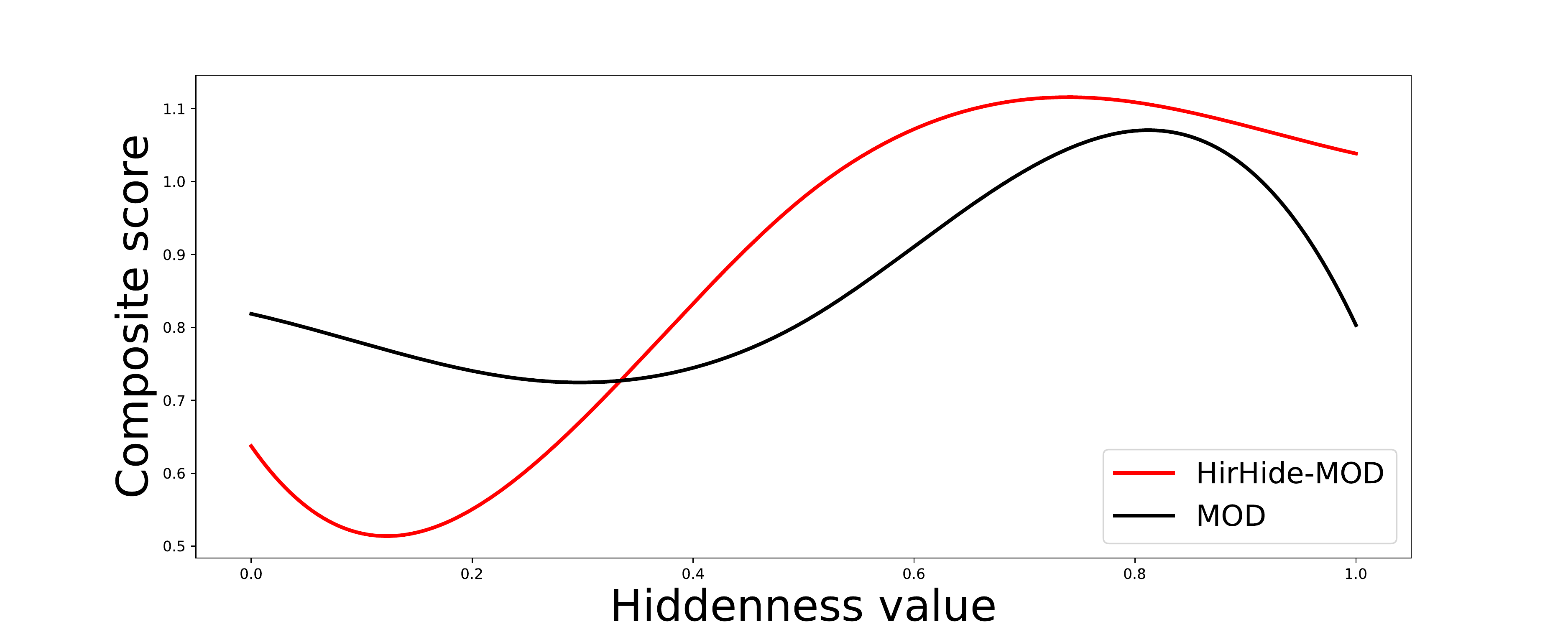}
\includegraphics[width=0.48\textwidth]{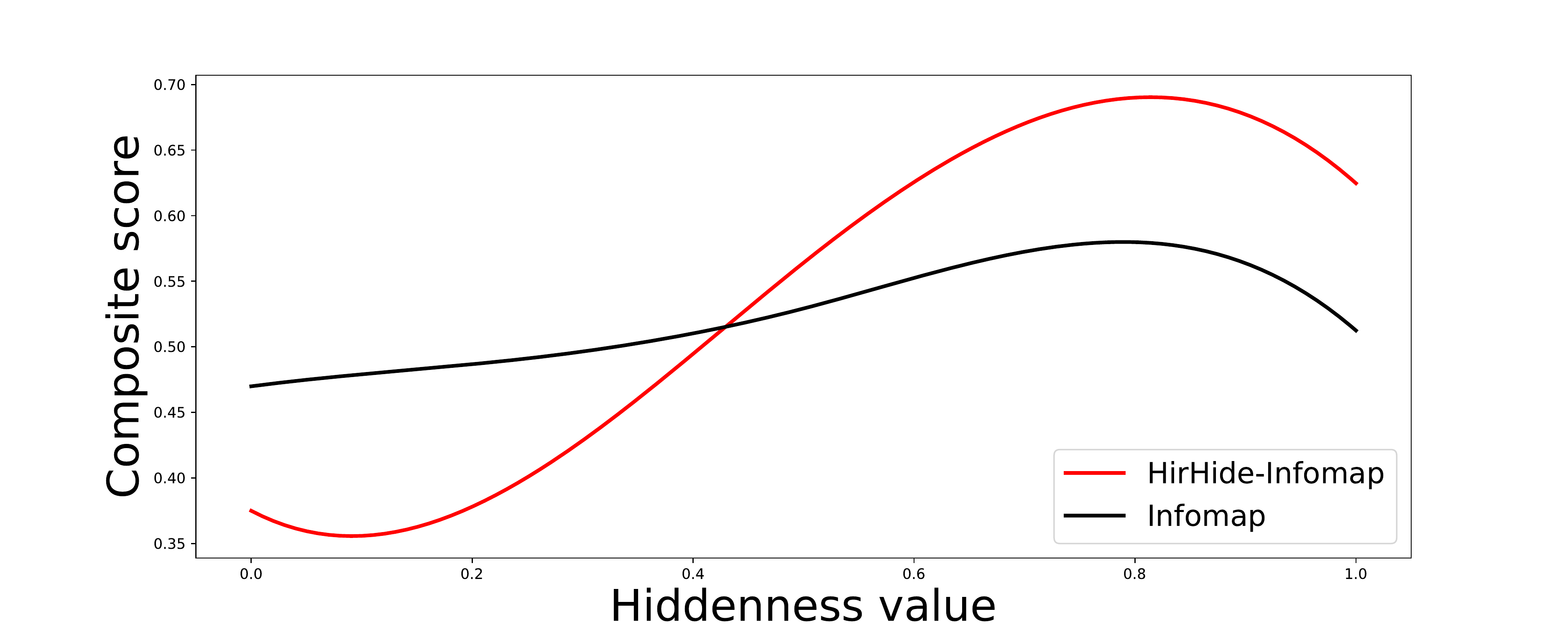}
\end{center}
\vspace{-1em}
\caption{The performance of different algorithms with the hiddenness value of communities increasing. HirHide-MOD detects more accurately than MOD on hidden communities with higher hiddenness values.
}
\label{hiddenness}
\end{figure*}

\subsection*{Experimental results on real-world data}
We compare the performance of standard algorithms with their HirHide versions and their HiCode versions in real-world networks. Fig. \ref{benchmark reuslts} shows the comparative performance of the ten algorithms using MIPS and CYC separately as the reference sets. The different bar clusters represent using different datasets ($Biogrid$, $Krogan\_core$ and $YeastNet$) as PPI networks. The bars on the bottom indicate the scores of maximum matching ratio (MMR) and the bars on top represent the scores of the recall. Higher bars represent better performance. In each bar, the specific score is shown if it is larger than 0.2. We can see that when benchmark algorithms are combined with HirHide, they achieve better MMR, recall and composite scores on most of the PPI datasets. As mentioned before, HirHide does not change the core of a standard algorithm but enables it to detect hierarchical hidden community structures. So the better performance demonstrates that detecting hierarchical hidden community structures helps detect complexes in PPI networks. Furthermore, the HirHide versions of the three benchmark algorithms perform better than their HiCode versions in all the datasets, which demonstrates that the hierarchical structure is necessary.\par

Fig. \ref{hiddenness} illustrates the performance of different algorithms with the hiddenness value of communities increasing. We show the results of HirHide-MOD versus MOD and HirHide-Infomap versus Infomap on $Biogrid$. The reference dataset is MIPS. A higher hiddenness value indicates a deeper hidden degree. In each sub-figure, we show the complexes detected by both algorithms.
Because there are hundreds of communities, we smooth the results to make them more concise. As illustrated in Fig. \ref{hiddenness}, HirHide-MOD and HirHide-Infomap show increasing advantage with the higher hiddenness value, indicating that HirHide has significant advantages in detecting hidden communities.

\begin{figure*}[!tpb]
\centering
\subfigure[]{\label{fig:a}\includegraphics[width=30mm]{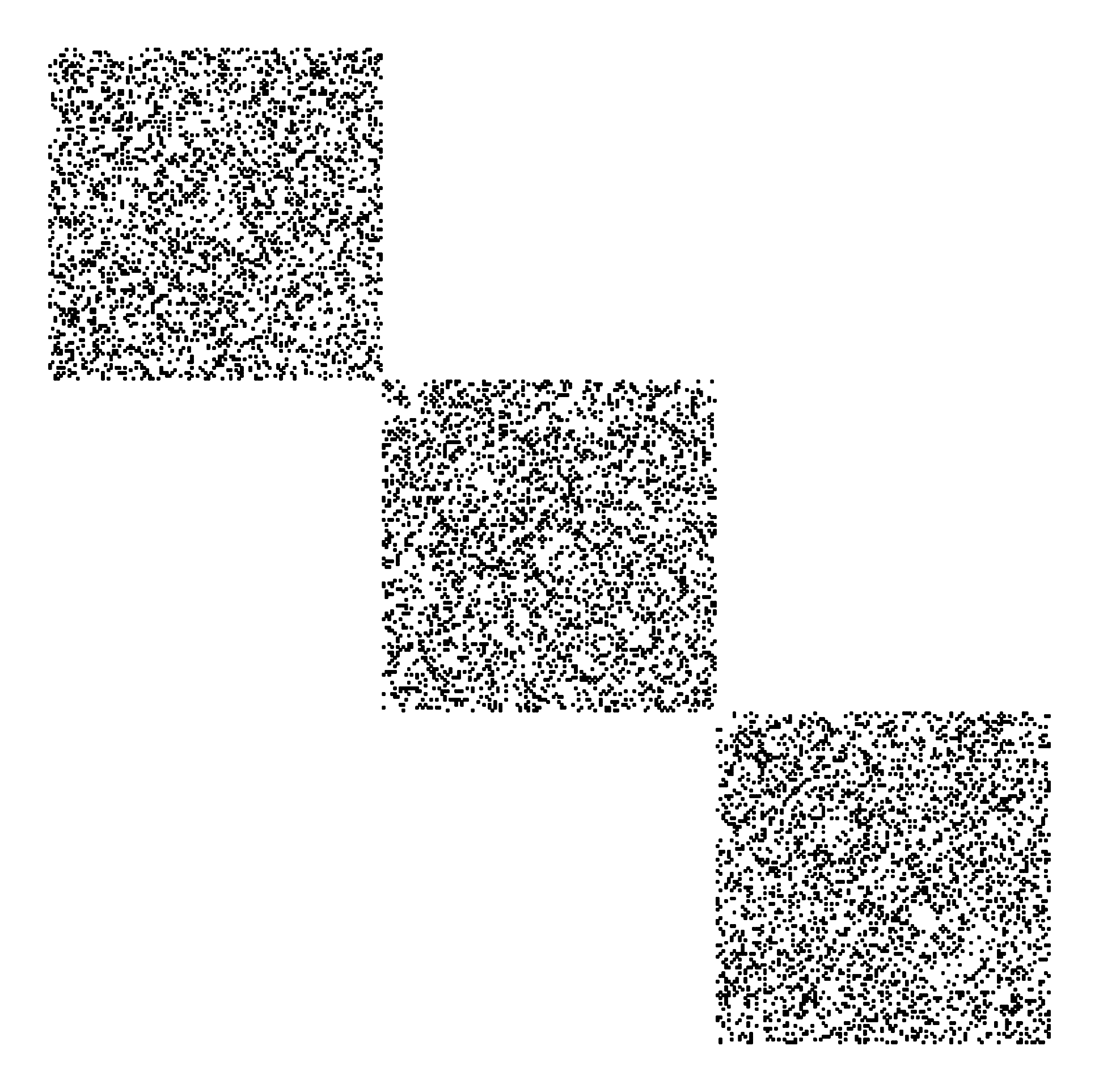}}
\subfigure[]{\label{fig:b}\includegraphics[width=30mm]{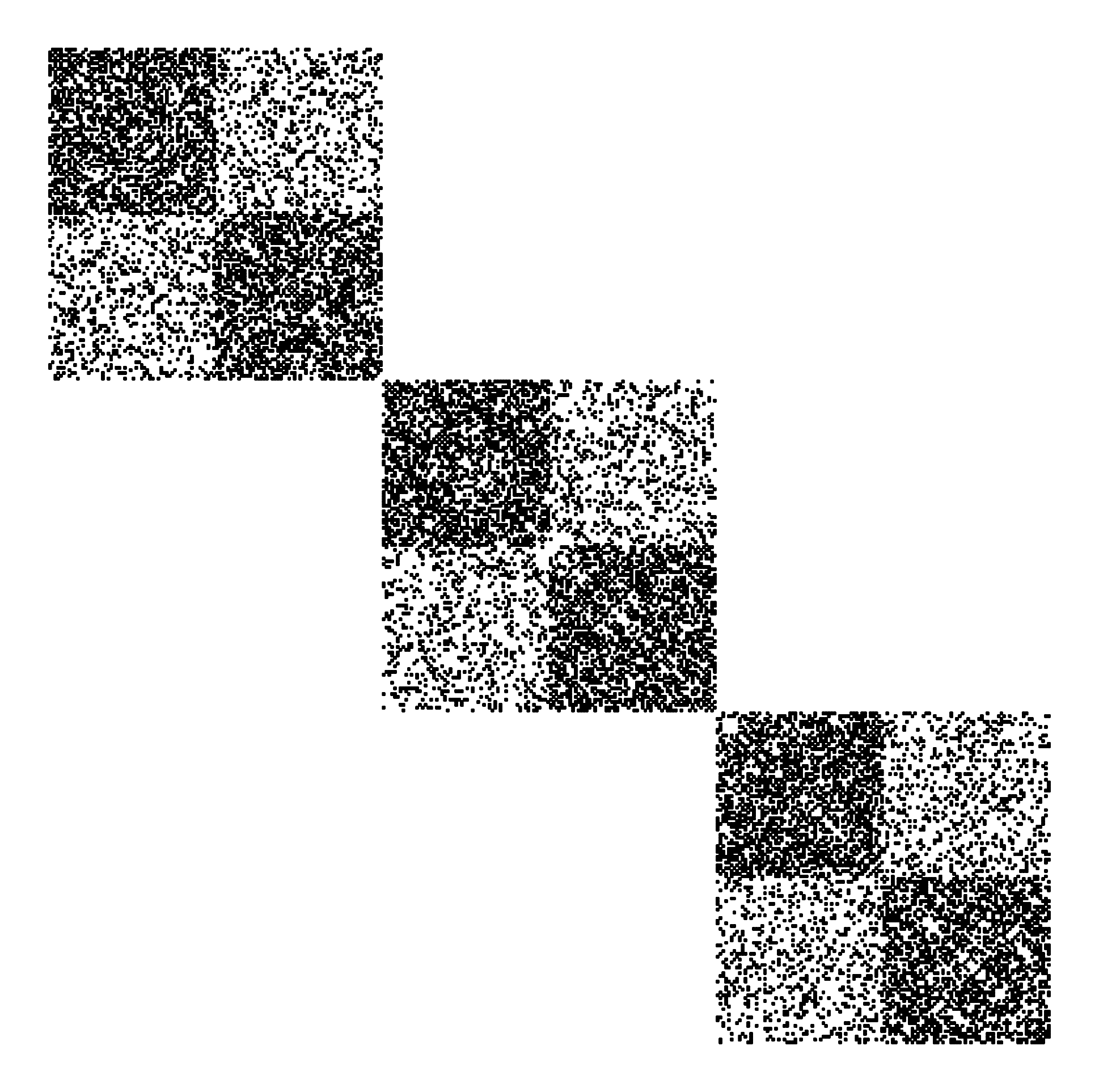}}
\subfigure[]{\label{fig:c}\includegraphics[width=30mm]{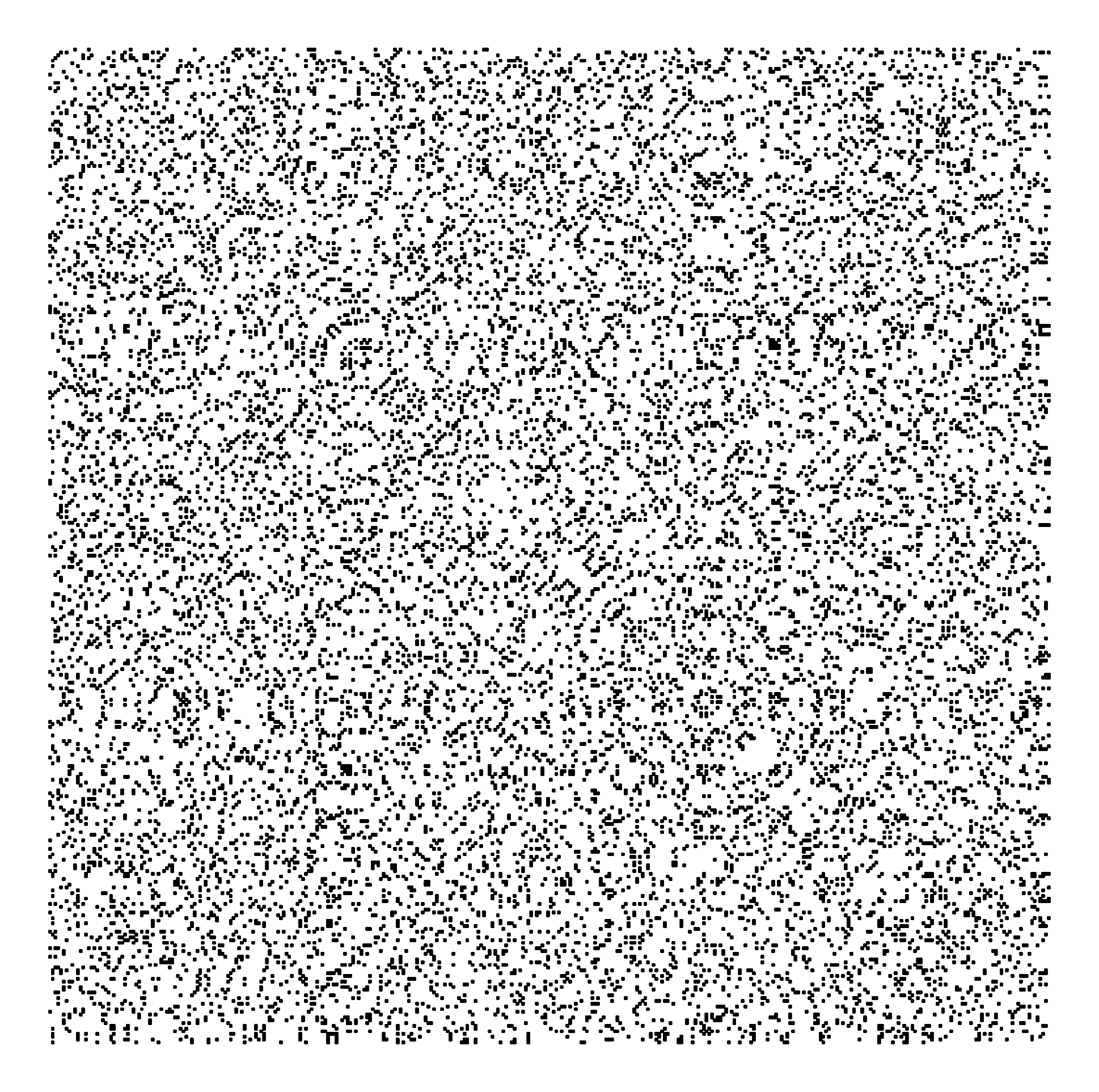}}
\subfigure[]{\label{fig:d}\includegraphics[width=30mm]{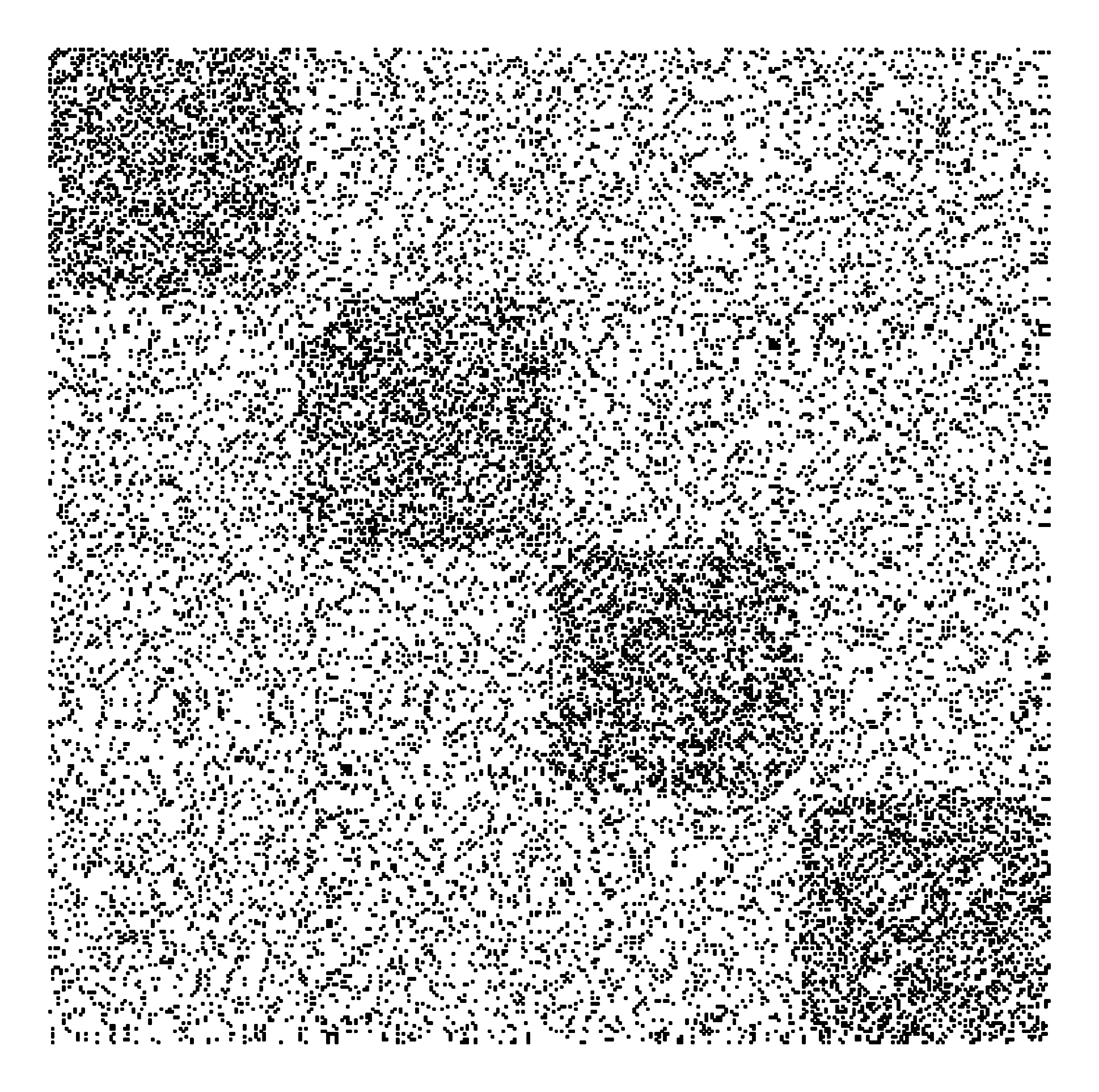}}
\subfigure[]{\label{fig:e}\includegraphics[width=30mm]{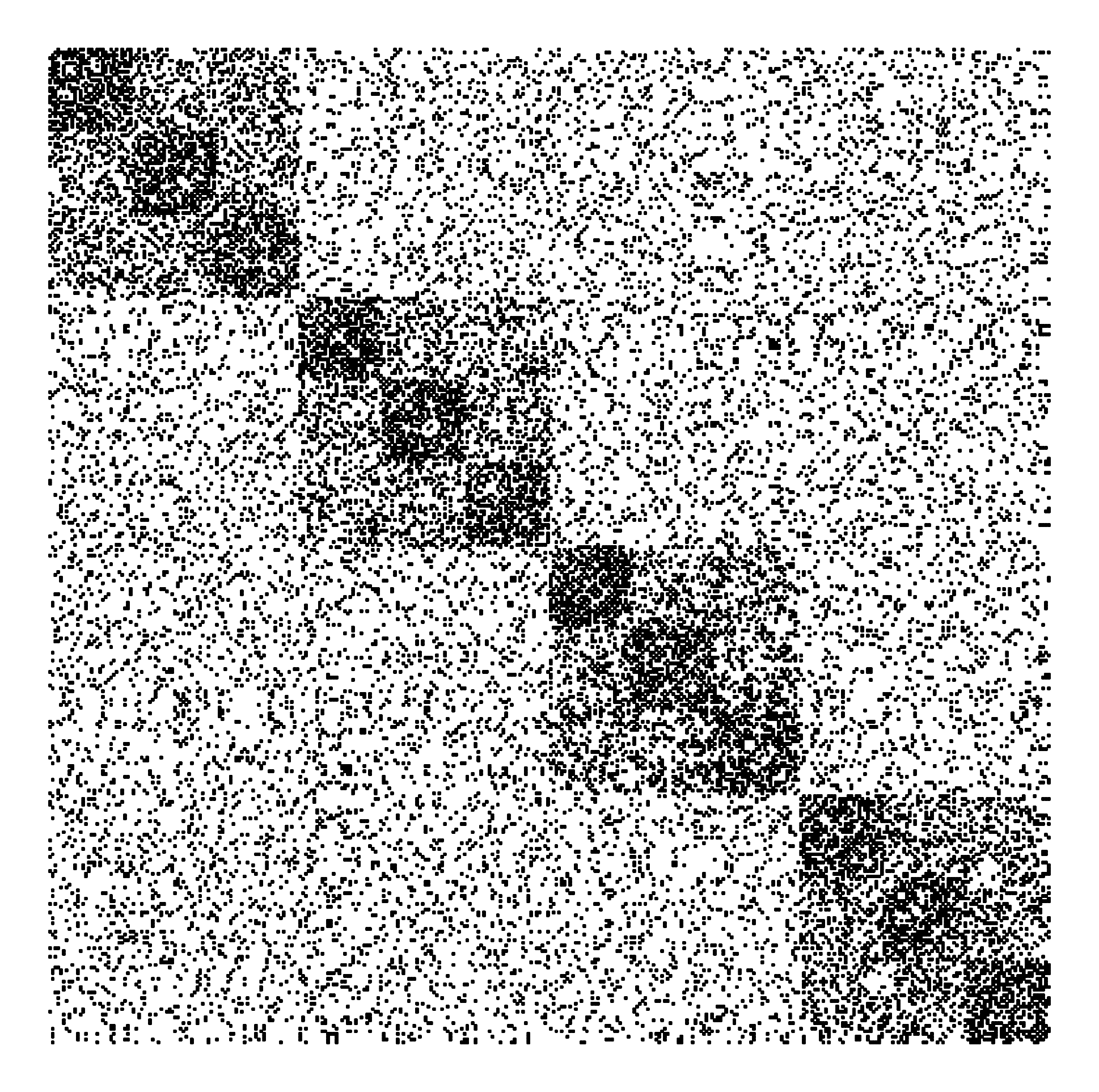}}
\caption{Constructing hierarchical hidden community structure. When constructing a community, it is treated as a random block model and edges are added within the community. (a) Constructing communities in root of the first layer. (b) Constructing communities in leaves of the first layer. (c) The graph after randomly scrambling the node number. (d) Constructing communities in root of the second layer. (e) Constructing communities in leaves of the second layer.}
\label{synthetic data}
\end{figure*}

\subsection*{Synthetic data and experimental results}
To some extent, the two evaluation metrics can solve the problem that the reference sets are incomplete, but they may be not persuasive enough. As a result, the comparison of synthetic networks is necessary. We build three synthetic networks, each of which contains two layers for hidden structure, wherein the first layer consists of strong communities and the second layer consists of relatively weak communities. In addition to the multi-layer feature, hierarchical structures are added. Specifically, in each layer, smaller and denser communities are added to make up the next hierarchical level.\par

The choice of edge probability follows a certain rule. According to the concept of hidden community, the interactions between proteins in the hidden layer are relatively sparse, so the edge probability is also lower than that of the first layer. When constructing the hierarchical structure, the edge probability of the leaf communities at the second level is slightly higher than the probability of the root communities, to further highlight the structure of sub-communities. The first network called $G_{360}$ contains 360 nodes. At the first layer, 360 nodes are divided into 3 communities, each with 120 nodes. According to the definition of the community, internal edges are added to the three communities with probability $P_{11}$ = 0.2, and no edge is added outside (Fig. \ref{synthetic data} (a)). Then the hierarchical structure is appended. The 120 nodes of each community are subdivided into two sub-communities. Six sub-communities make up the second hierarchical level, each of which contains 60 nodes. Edges are added to the inside of the six sub-communities with probability $P_{12}$ = 0.3 (Fig. \ref{synthetic data} (b)). Then the first layer is successfully built.\par

According to the idea of hidden structure, before building the second layer, the node numbers in the first layer are randomly scrambled so that the previous community structure is evenly distributed in the adjacency matrix (Fig. \ref{synthetic data} (c)).\par
The second layer is constructed on the randomly scrambled graph. In the second layer, 360 nodes are divided into four communities, each of which has 90 nodes. Internal edges are added in these communities with a probability of $P_{21}$ = 0.15 (Fig. \ref{synthetic data} (d)). Then we add a hierarchical structure, and each community is divided into three small sub-communities, each of which has 30 nodes. Edges are added inside these sub-communities with a probability of $P_{22}$ = 0.25 (Fig. \ref{synthetic data} (e)). \par
Similarly, we also construct a network with 2000 nodes and a network with 3000 nodes, Called $G_{2000}$ and $G_{3000}$. \par

\begin{table}[!t]
\centering
\caption{The performance of HirHide-MOD (H-MOD) and MOD on synthetic data.}
\footnotesize
\vspace{2pt}
\setlength{\tabcolsep}{1.0mm}
{\begin{tabular}{@{}lllllll@{}}
\toprule \textbf{Truth}&\multicolumn{2}{c}{\textbf{$G_{360}$}}&\multicolumn{2}{c}{\textbf{$G_{2000}$}}&\multicolumn{2}{c}{\textbf{$G_{3000}$}}\\ 
&H-MOD&MOD&H-MOD&MOD&H-MOD&MOD\\ \midrule
\textbf{$L_{11}$}&1.000&1.000 &0.999 &0.980 &1.000 &0.999 \\
\textbf{$L_{12}$}&0.997&0.857 &0.686&0.681 &0.859 &0.690 \\
\textbf{$L_{21}$}& 1.000& \textbf{0.337} &1.000 &\textbf{0.100} &1.000 &\textbf{0.079} \\
\textbf{$L_{22}$}&0.606&\textbf{0.215 } &0.474 &\textbf{0.109} &0.721 &\textbf{0.080} \\ \toprule
\end{tabular}}
\label{result_synthetic}
\end{table}

Because MOD has the best performance in the three benchmark algorithms, We compare HirHide-MOD with MOD on our synthetic data. The experimental results are shown in Table \ref{result_synthetic}. $L_{11}$, $L_{12}$ represent the dominant community layer and $L_{21}$, $L_{22}$ represent the hidden community layer. As illustrated in Table \ref{result_synthetic}, HirHide-MOD has a slight advantage over MOD on $L_{11}$ and $L_{12}$, which means HirHide doesn't reduce the performance of traditional algorithms in networks without hidden structure. Moreover, HirHide-MOD has significant advantage over MOD on $L_{21}$ and $L_{22}$. And the scores of MOD on $L_{21}$ and $L_{22}$ are extremely low while the scores of HirHide-MOD on $L_{21}$ and $L_{22}$ are normal and high. Consequently, we can conclude that MOD can not detect the hidden community layer. But after it is combined with the HirHide framework, HirHide-MOD can detect the hidden community layer well. These results are consistent with the results in real-world networks.\par

\begin{table}[!t]
\centering
\caption{Results of weakening communities in roots or leaves. \textit{Weakening Roots} means that weakening communities in roots and \textit{Weakening Leaves} means that weakening communities in leaves.}
\footnotesize
\vspace{2pt}
\setlength{\tabcolsep}{3.0mm}
{\begin{tabular}{@{}ccc@{}}
\toprule\textbf{Truth}&\textbf{Weakening Roots}&\textbf{Weakening Leaves}\\\midrule
\textbf{$L_{11}$}&\textbf{1.000}&0.913\\
\textbf{$L_{12}$}&\textbf{0.997}&0.691\\
\textbf{$L_{21}$}&\textbf{1.000}&0.911\\
\textbf{$L_{22}$}&\textbf{0.606}&0.485\\ \toprule
\end{tabular}}
\label{weaken}
\end{table}

\begin{table}[ht]
\centering
\caption{Results of grabbing sub-communities on original graph or weakened graph. }
\footnotesize
\vspace{2pt}
\setlength{\tabcolsep}{1.2mm}
{\begin{tabular}{@{}lllll@{}}
\toprule \textbf{Graph}&\multicolumn{2}{c}{\textbf{$L_{12}$}}&\multicolumn{2}{c}{\textbf{$L_{22}$}}\\ 
&Original &Weakened &Original&Weakened\\\midrule
\textbf{$G_{2000}$}&0.686&\textbf{0.707}&0.474&\textbf{0.479}\\
\textbf{$G_{3000}$}&\textbf{0.859}&0.826&\textbf{0.721}&0.717\\ \toprule
\end{tabular}}
\label{original}
\end{table}

\begin{figure}[!tpb]
\centering
\subfigure[]{\label{fig:a}\includegraphics[width=40mm]{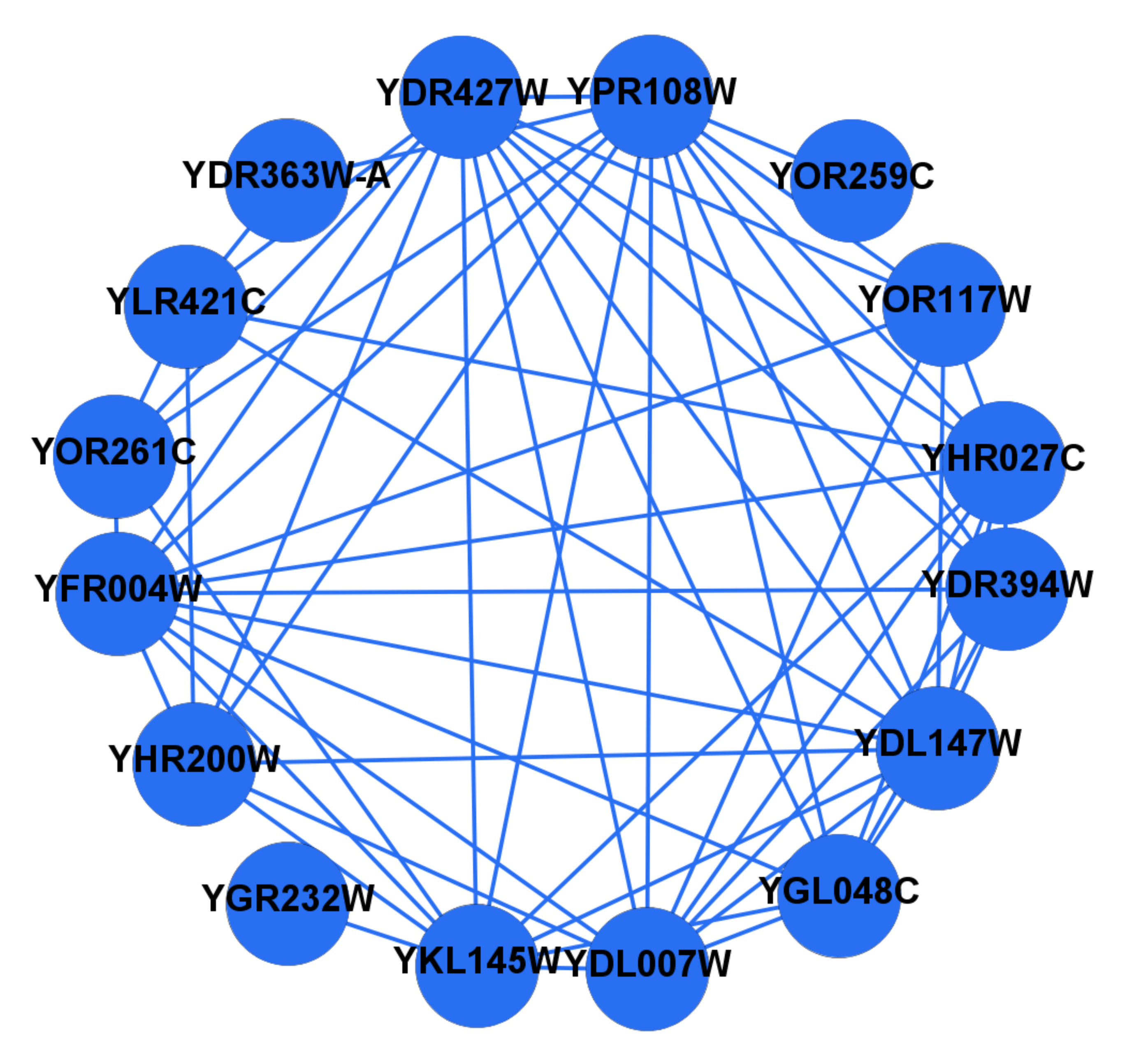}}
\subfigure[]{\label{fig:b}\includegraphics[width=40mm]{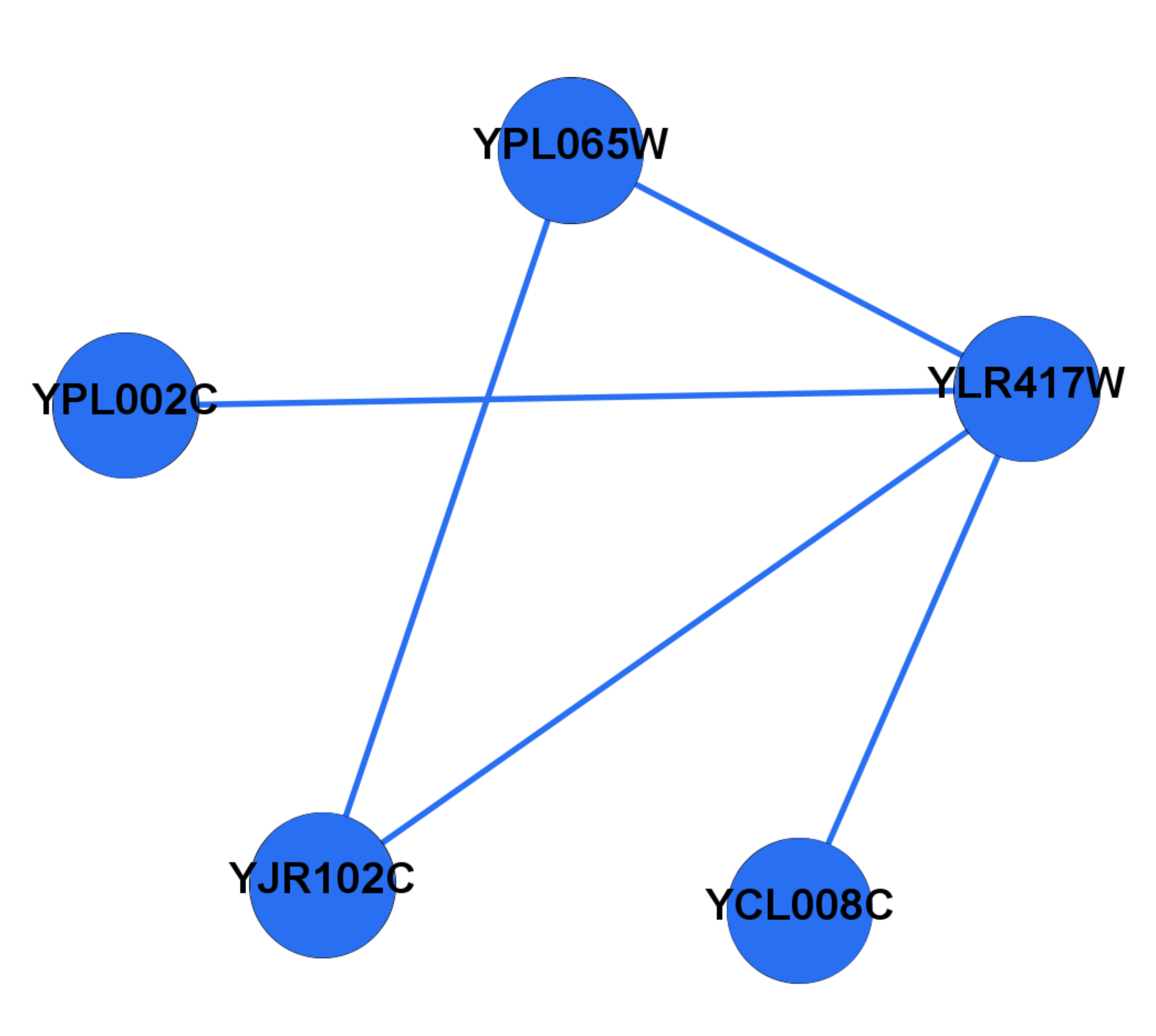}}
\caption{Two examples of predicted protein complexes identified by HirHide. \label{examples} }
\end{figure}

\subsection*{Confirmatory experiments}
In the HirHide framework, the strong community structure detected by the base algorithm needs to be weakened. After combining the concept of hierarchical structure, there is a choice of weakening communities at the top level of the hierarchical structure, or weakening communities at the lowest level of the hierarchical structure. To determine which level we should choose, we experiment on the synthetic network $G_{3000}$. Because the synthetic data is complete, we have chosen the F1 score as the evaluation metric. As illustrated in Table \ref{weaken}, weakening communities in roots can make HirHide have a better performance, especially in level 2.\par

In HirHide, there are two options of grabbing sub-communities in a layer: to grab sub-communities on the original graph or to grab sub-communities on the reduced graph after weakening other layers. We still use the F1 score to evaluate the performance. As illustrated in Table \ref{original}, in graph $G_{3000}$, grabbing the sub-communities on the original graph has a slight advantage. But in graph $G_{2000}$, we have an opposite result. Consequently, overall which graph is chosen does not make a big difference.\par

\subsection*{Screening predicted complexes by emerging patterns}
A HirHide-combined algorithm can detect additional hierarchical hidden communities that can not be detected without HirHide. Some of them fit the characteristics of protein complexes with sparse internal connections. So they can serve as predicted protein complexes. Here, the predicted protein complexes mean these protein complexes do not appear in the reference sets and they are potential to be protein complexes. \par

However, detecting these hierarchical hidden communities are solely based on density property in PPI networks. Not all of them are reliable enough. Emerging patterns (EPs) are conjunctive patterns that contrast sharply between different classes of data, which contain more informative properties such as degree statistics, clustering coefficient, topological coefficients and eigenvalues of a sub-graph. Recently, EPs are exploited to address the complex prediction problem \cite{liu2016using}. We screen more reliable predicted protein complexes based on this EP-based method. A feature vector is first constructed to describe the key properties of the reference protein complexes as well as those of random non-complexes communities. Then to discover EPs by contrasting feature vectors of reference protein complexes and random non-complexes communities. Next, the discovered EPs are used to discover potential complexes. For each of the complexes predicted by HirHide, if it is similar to complexes discovered by EPs, it is considered as a reliably predicted complex. Fig. \ref{examples} illustrates two examples of the predicted complexes.
\par

\section{Discussion}

The main contribution of this work is to propose a new method called HirHide and apply HirHide for protein complex detection and prediction. HirHide serves as a meta-method that can be combined with existing standard community detection algorithms and enable them to discover hierarchical hidden communities. We improve the definition of the necessity of detecting hidden structures and why detecting hierarchical hidden structures is more helpful for complex detection in PPI networks. We redefine the hiddenness value to fit large-scale and complicated networks. And we have compared standard detection algorithms with their HirHide versions in synthetic data and real-world PPI networks. Experimental results illustrate that the performance of standard algorithms is boosted when combined with HirHide. In networks without hidden structure, HirHide doesn't reduce their performance.\par

How to combine the concept of hierarchical structure with hidden structure is a key issue in our work. We verify which level of the hierarchical structure is selected when iterating in HirHide, and conclude that weakening communities at the root level can achieve the best weakening effectiveness and there is almost no difference to grab sub-communities on the original graph or the reduced graph. Finally, we treat additional hierarchical hidden communities as preliminary prediction of undiscovered protein complexes. And further screening these predicted complexes by emerging patterns. These steps guarantee that the predicted complexes are reliable.\par

In this work, we applied HirHide to detect protein complexes in PPI networks. Actually, HirHide can be used for most community detection questions because of the flexibility of the base algorithms. HirHide is also suitable for questions related to other large-scale biological data such as neural networks and gene regulatory networks. Some of these biological networks are too complex for common community detection algorithms to produce a positive performance. Most of the time, a better choice is to choose a corresponding algorithm based on the characteristics of a certain biological network. Under these circumstances, HirHide has clear advantages because of its flexibility. At the same time, according to the natural hierarchical structure of cells, organelle, intracellular compound \textit{etc.}, hierarchical hidden community structure is in line with the characteristics of the data itself, thus helping researchers to study biological interactions more deeply.\par

\bibliographystyle{ieee.bst}
\bibliography{document.bib}

\end{document}